
\magnification=\magstep1
\overfullrule=0pt
\hsize=15.4truecm
{\nopagenumbers \line{\hfil UQAM-PHE-95/10}
\vskip.51cm
\centerline{\bf BOUNDS ON SECOND GENERATION SCALAR LEPTOQUARKS}
\centerline{\bf FROM THE ANOMALOUS MAGNETIC MOMENT OF THE MUON}
\vskip3cm
\centerline{G. COUTURE AND H. K\"ONIG
\footnote*{couture@mercure.phy.uqam.ca, konig@osiris.phy.uqam.ca}}
\centerline{D\'epartement de Physique}
\centerline{Universit\'e du Qu\'ebec \`a Montr\'eal}
\centerline{C.P. 8888, Succ. Centre Ville, Montr\'eal}
\centerline{Qu\'ebec, Canada H3C 3P8}
\vskip2cm
\centerline{\bf ABSTRACT}\vskip.51cm\indent
We calculate the contribution
of second generation scalar leptoquarks to the anomalous
magnetic moment of the muon (AMMM). In the near future,
E-821 at Brookhaven
will reduce the experimental error on this parameter to
$\Delta a_\mu^{\rm exp}<4\times 10^{-10}$, an improvement of $20$\ over its
current value.
With this new
experimental limit we obtain a lower mass limit of
$m_{\Phi_L}>186$\ GeV for the second generation scalar leptoquark,
when its Yukawa-like coupling $\lambda_{\Phi_L}$\
to quarks and leptons is taken to be of the order of
the electroweak coupling $g_2$.
\vskip.51cm
\centerline{ JULY 1995}
\vfill\break}
\pageno=1
Recently there has been a lot of interest in leptoquarks theoretically
[1--4] as well as experimentally [5--8] (and references within).
Leptoquarks are quite common in extensions of the Standard Model.
They are $SU(3)_C$\ triplets and can occur
as vectors or scalars. Since vector quarks are of more difficult
nature in a low-energy theory we restrict ourselves in this
Brief Report to scalar leptoquarks, which are $SU(2)$\ doublets.
\hfill\break\indent
There exist already severe constraints on the characteristics
of leptoquarks. Flavour changing neutral current (FCNC) forces
the leptoquarks to be flavour diagonal; that is we have to
suppose that they couple to only a single generation of leptons
and quarks [9]. On the other hand from pseudoscalar mesons
($\pi,\ K$\ and $D$) leptonic decays we also have to conclude
that leptoquarks couple only chirally; that is they couple either
to left-handed or to right handed quarks only, but not to both
at the same time [10].
\hfill\break\indent
The most stringent lower mass bound for the first generation
leptoquarks is given by HERA experiments [8] $m_{\Phi_L}>145$\ GeV,
where the Yukawa-like coupling of the leptoquarks to electrons
and up quarks was supposed to be of electromagnetic strength
$\lambda_{\Phi_L}=e$. In a recent report the CDF collaboration
[7 the latter] presented for the second generation a lower
mass bound of $m_{\Phi_L}>135$\ GeV if the branching ratio into muon and quark
is taken to be $100$\% and $m_{\Phi_L}>95$\ GeV for
a branching ratio of $50$\%\footnote*{In a recent report the
D0 obtains $m_{\Phi_L}>111$\ GeV for a branching ratio
of $100$\% and $m_{\Phi_L}>89$\ GeV for a branching ratio of $50$\%[11]}.
 In [4] the authors obtain
via the leptonic partial widths of the Z boson a lower
mass limit of about $680$\ GeV for a left-type leptoquark
and of about $280$\ GeV for a right-type leptoquarks when the
top mass is taken to be $180$\ GeV.
\hfill\break\indent
In this Brief Report we consider the contribution of
second generation scalar leptoquarks to the AMMM. The QED contribution has
been calculated to eighth order and estimated to tenth order[12].
The Standard Model contributions to the AMMM
are also very well know: the W-boson contribution is the most important
($\Delta a^W_\mu\sim 40\times 10^{-10}$), the Z contribution interferes
destructively ($\Delta a^Z_\mu\sim -20\times 10^{-10}$), and a heavy neutral
Higgs contribution is totally negligible[13]. The charged Higgs contribution is
also very small for a mass of a few GeV's or more [14], while the
supersymmetric contributions can be large in certain mass domains [15].
Recently, two groups have performed 2-loop
calculations in the context of the SM: the authors
found that those
contributions can be as large as 10\%-12\% of the 1-loop SM
contribution [16].
In the near future, E-821 at Brookhaven will reduce
the experimental error on the AMMM by a factor
of $20$\ [17] thus leading to an experimental error
of $\Delta a_\mu^{\rm exp}<4\times 10^{-10}$. The main goal is to see the
weak contributions but it is clear that such a measurement
will lead to severe constraints for models beyond the
SM.
\hfill\break\indent
The diagrams leading to the AMMM are given in Fig.1 and
the Lagrangian that describes the coupling of the leptoquarks
to photon and to the strange quark and muon is given by [4,18]
$${\cal L}=+e e^q_{\Phi_L}(k_1+k_2)^\mu \Phi_L^\dagger\Phi_L
A^\mu+\lambda_{\Phi_L}\overline\mu\lbrack g_L^qP_L+g_R^qP_R\rbrack
q\Phi\eqno(1)$$
with $e^q_{\Phi_L}=-2/3,-5/3$ if $q=s,c$\ respectevely.
The chiral coupling of the leptoquarks forces us to
consider $g_L^q=1,\ g_R^q=0$\ (left-type leptoquark)
or $g_L^q=0,\ g_R^q=1$\ (right-type leptoquark).
The calculation of the first and second diagram shown
in Fig.1 and the last two self energy diagrams lead to the following
results:
$$\eqalignno{iM_1=&+{{e e_q\lambda_{\Phi_L}^2}\over{(4\pi)^2}}
g_a^{q2}\int\limits_0^1 d\alpha_1\int\limits_0^{1-\alpha_1}
d\alpha_2\overline u_{p_2}\biggl\lbrace\lbrack {1\over\epsilon}
-1-\gamma+\log(4\pi\mu^2)-\log G(m_{\Phi_L}^2,m_q^2)\rbrack P_a\gamma_\mu
\cr&+P_a\lbrack (\rlap/p_2-\rlap/\tilde p)\gamma_\mu(\rlap/p_1-
\rlap/\tilde p)+m_q^2\gamma_\mu\rbrack{1\over{G(m_{\Phi_L}^2,m_q^2) }}
\biggr\rbrace u_{p_1}\epsilon^{\ast\mu}&(2)\cr
iM_2=&+{{e e^q_{\Phi_L}\lambda_{\Phi_L}^2}\over{(4\pi)^2}}
g_a^{q2}\int\limits_0^1 d\alpha_1\int\limits_0^{1-\alpha_1}
d\alpha_2\overline u_{p_2}\biggl\lbrace\lbrack {1\over\epsilon}
-\gamma+\log(4\pi\mu^2)-\log G(m_q^2,m_{\Phi_L}^2)\rbrack P_a\gamma_\mu
\cr&+P_a\lbrack\rlap/\tilde p(p_1+p_2-2\tilde p)^\mu
{1\over{G(m_q^2,m_{\Phi_L}^2) }}\biggr\rbrace u_{p_1}\epsilon^{\ast\mu}\cr
iM_{SE}=&+{{e\lambda_{\Phi_L}^2}\over{(4\pi)^2}}g_a^{q2}\int\limits_0^1
d\alpha_1\alpha_1\overline u_{p_2}\biggl\lbrace\lbrack {1\over\epsilon}
-\gamma+\log(4\pi\mu^2)-\log H(m_q^2,m_{\Phi_L}^2)\rbrack P_a\gamma_\mu
\biggr\rbrace u_{p_1}\epsilon^{\ast\mu}\cr
G(m_i^2&,m_j^2) =~m_i^2-(m_i^2-m_j^2)(\alpha_1+
\alpha_2)-m_\mu^2(\alpha_1+\alpha_2)(1-\alpha_1-\alpha_2)
-q^2\alpha_1\alpha_2\cr
H(m_i^2&,m_j^2) =~m_i^2-(m_i^2-m_j^2)\alpha_1-
m_\mu^2\alpha_1(1-\alpha_1)\cr
\tilde p=&p_1\alpha_1+p_2\alpha_2\cr}$$
with $e_q=-1/3,+2/3$\ for $q=s,u$\ respectively
and $a=L,R$. $q^2=(p_1-p_2)^2=0$\ for a real photon.
 After the summation of all diagrams the divergencies
cancel out. To obtain the AMMM we have to make use of gauge invariance
$q^\mu\epsilon^{\ast\mu}=0$\ that is $p_1^\mu\epsilon^{\ast\mu}=
p_2^\mu\epsilon^{\ast\mu}$, of the Dirac equation $\overline u_{p_2}
P_L\rlap/ p_2 u_{p_1}=m_\mu u_{p_2}P_R u_{p_1}$\ ($\overline u_{p_2}
P_L\rlap/ p_1u_{p_1}=m_\mu u_{p_2}P_Lu_{p_1}$) and perform a Gordon
decomposition $2\overline u_{p_2}p_1^\mu u_{p_1}=
\overline u_{p_2}(i\sigma_{\mu\nu}q^\nu+2 m_\mu\gamma_\mu)u_{p_1}$.
As a finite result we obtain a term
proportional to the $\gamma_\mu$\ term, which has to be renormalised
by adding a counter term [4] and a term proportional to
$\sigma_{\mu\nu}q^\nu$\ from which the
AMMM $a_\mu$\ can be extracted by
$V_\mu={e\over{2m_\mu}}a_\mu\overline u_{p_2}i\sigma_{\mu\nu}q^\nu u_{p_1}$.
After some calculation the contribution of the second
generation scalar leptoquark to the AMMM is given by
$$\eqalignno{
\Delta a_\mu^{\Phi_L}=&+{{\lambda_{\Phi_L}^2}\over{(4\pi)^2}}
g_a^{q2}m_\mu^2\int\limits_0^1d\alpha_1\alpha_1^2(1-\alpha_1)
\lbrack {{e^q_{\Phi_L}} \over
{H(m_q^2,m_{\Phi_L}^2)}}-{{e_q}\over{H(m_{\Phi_L}^2,m_q^2)}}\rbrack&(3)\cr
=&+{{\lambda_{\Phi_L}^2}\over{(4\pi)^2}}g_a^{q2}{1\over 6}
(e^q_{\Phi_L}-2e_q)({{m_\mu}\over{m_{\Phi_L}}})^2
\cr}$$
Eq.(3) is exact while the last result makes use of
$m_{\Phi_L}^2\gg m_s^2,\ m_\mu^2$.
As a result we have that for $q=s$\ the first order
in the expansion of $1/m_{\Phi_L}^2$\ is identical to $0$\
($e^s_{\Phi_L}-2e_s\equiv 0$), whereas for $q=c$\ we have
$e^c_{\Phi_L}-2e_c=-3$.
Parametrizing the Yuakawa-like leptoquark coupling with electoweak strength
$\lambda_{\Phi_L}^2=g_2^2k$\
($k=1-10$) [4] we have $\vert\Delta a_\mu^{\Phi_L}\vert=
1.38\times 10^{-5}k({\rm GeV}/m_{\Phi_L})^2$\
when the charm quark is taken within the loop. Combining this result with the
expected experimental error on
the AMMM ($\Delta a_\mu^{\rm exp}=4\times 10^{-10}$)\ we obtain
a lower mass limit of a second generation scalar lepton quark
(left-type or right-type) of $m_{\Phi_L}>186$\ GeV
for $k=1$ and $m_{\Phi_L}>588$\ GeV for $k=10$. These bounds
compare very well with direct searches
at current accelerators.
\hfill\break\vskip.1cm\noindent
{\bf ACKNOWLEDGMENTS}\vskip.12cm
\noindent
We want to thank Mike A. Doncheski for clarifying discussions
and pointing out ref.11.
This work was partially funded by the N.S.E.R.C. of
Canada and les Fonds F.C.A.R. du Qu\'ebec.
\hfill\break\vskip.12cm\noindent
{\bf REFERENCES}\vskip.12cm
\item{[\ 1]}J. Bl\"umlein and R. R\"uckl, Phys.Lett.{\bf B304}
(1993)337, J.E. Cieza Montavlo and O.J. \'Eboli, Phys.Rev.
{\bf D47}(1993)837, J. Bl\"umlein, E. Boos and A. Pukhov,
Mod.Phys.Lett.{\bf A9}(1994)3007; see also the recent review
{\it New Particles and Interactions} by A. Djouadi, J. Ng, T.G. Rizzo,
{\it et al},
SLAC-PUB-95-6772, Mar 1995, 64pp,
to appear as a chapter in {\bf Electroweak Symmetry Breaking and Beyond the
Standard Model}, edited by T. Barklow, S. Dawson, H.E. Haber and S.
Siegrist, World Scientific: hep-ph/9504210.
\item{[\ 2]}G. B\'elanger, D. London and H. Nadeau, Phys.Rev.{\bf
D49}(1994)3140, M.A. Doncheski and S. Godfrey, Phys.Rev.{\bf
D49}(1994)6220, J.K. Mizukoshi, O.J. \'Eboli and M.C. Gonzal\'ez-
Gracia, CERN preprint, CERN-TH-7508-94.
\item{[\ 3]}M. Leurer, Phys.Rev.Lett.{\bf 71}(1993)1324,
S. Davidson, D. Bailey and B.A. Campbell, Z.Phys.{\bf C61}(1994)613.
\item{[\ 4]}G. Bhattacharyya, J. Ellis and K. Sridhar, Phys.Lett.{\bf
B336}(1994)100, Erratum-ibid{\bf B338}(1994)522.
\item{[\ 5]}DELPHI Collaboration, P. Abreu et al., Phys.Lett.{\bf B316}
(1993)620; L3 Collaboration, B. Adeva et al., Phys.Lett.{\bf B261}
(1991)169; OPAL Collaboration, G. Alexander et al., Phys.Lett.{\bf B263}
(1991)123.
\item{[\ 6]}D0 Collaboration, S. Abachi et al., Phys.Rev.Lett.{\bf 72}
(1994)965.
\item{[\ 7]}CDF Collaboration, F. Abe et al., Phys.Rev. {\bf D48}(1993)
3939, FERMILAB-PUB-95-050-E.
\item{[\ 8]}H1 Collaboration, I. Abt et al., Nucl.Phys.{\bf 396}
(1993)3, Z.Phys.{\bf C64}(1994)545, ZEUS Collaboration, M. Derrick
et al., Phys.Lett.{\bf B396}(1993)173.
\item{[\ 9]}W. Buchm\"uller and D. Wyler, Phys.Lett.{\bf B177}(1986)377,
 J.C. Pati and A. Salam, Phys.Rev.{\bf D10}(1974)275.
\item{[10]}O. Shankar, Nucl.Phys.{\bf B204}(1982)375; {\bf206}(1982)253.
\item{[11]}D0 Collaboration, S. Abachi et al., hep-ex/9507002.
\item{[12]}T. Kinoshita, B. Nizic, Y. Okamoto, Phys.Rev.{\bf D41}(1990)593.
\item{[13]}See for example, R. Jackiw, S. Weinberg, Phys.Rev.{\bf
D5}(1972)2473;
G.Altarelli, N. Cabibbo, L. Maiani, Phys.Lett.{\bf B40}(1972)415;
W.A. Bardeen, R. Gastmans, B.E. Lautrup, Nucl.Phys.{\bf B46}(1972)319.
\item{[14]}J.A. Grifols, R. Pascual, Phys.Rev.{\bf D21}(1980)2672.
\item{[15]}See for example, H. K\"onig, Z.Phys.{\bf C52}(1991)159,
Mod.Phys.Lett.{\bf A7}(1992)279 and references
therein.
\item{[16]}A. Czarnecki, B. Krause, W.J. Marciano, preprint TTP-95-19:
hep-ph/9506256; S. Peris, M. Perrottet, E. de
Rafael, Preprint CPT-95/P3202, CERN-TH/95-141:
hep-ph/9505405.
\item{[17]}V.M. Hughes, AIP Conference Proceedings, Intersections
between Part. and Nucl.Phys., Lake Louise, Canada 1986; J. Bailey {\it et al},
Nucl. Phys. {\bf B150}(1979)1.
\item{[18]}See for example W. Buchm\"uller, R. R\"uckl, D. Wyler,
Phys.Lett.{\bf B191}(1987)442.
\hfill\break\vskip.12cm\noindent
{\bf FIGURE CAPTIONS}\vskip.12cm
\item{Fig.1}The penguin and self energy diagrams with
the second generation scalar leptonquark and
quarks contributing to the AMMM. The $q$\ within the
loop denotes an up or strange quark, whereas the $q$\ at
the photon denotes its momenta $q=p_1-p_2$.
\vfill\break
\end